\documentclass[sigconf]{acmart}
\usepackage{soul}
\usepackage{multirow}
\usepackage{float}

\AtBeginDocument{%
  \providecommand\BibTeX{{%
    \normalfont B\kern-0.5em{\scshape i\kern-0.25em b}\kern-0.8em\TeX}}}

\copyrightyear{2023}
\acmYear{2023}
\setcopyright{acmlicensed}\acmConference[WWW '23]{Proceedings of the ACM Web Conference 2023}{May 1--5, 2023}{Austin, TX, USA}
\acmBooktitle{Proceedings of the ACM Web Conference 2023 (WWW '23), May 1--5, 2023, Austin, TX, USA}
\acmPrice{15.00}
\acmDOI{10.1145/3543507.3583321}
\acmISBN{978-1-4503-9416-1/23/04}

\begin{document}

\title{A Counterfactual Collaborative Session-based \\Recommender System}


\author{Wenzhuo Song}
\affiliation{%
  \institution{Northeast Normal University}
  \city{Changchun}
  \country{China}}
\email{wzsong@nenu.edu.cn}

\author{Shoujin Wang}
\affiliation{
  \institution{University of Technology Sydney}
  \city{Sydney}
  \country{Australia}}
\email{shoujin.wang@uts.edu.au}

\author{Yan Wang}
\affiliation{
  \institution{Macquarie University}
  \city{Sydney}
  \country{Australia}}
\email{yan.wang@mq.edu.au}

\author{Kunpeng Liu}
\affiliation{
  \institution{Portland State University}
  \city{Portland}
  \country{United States}}
\email{kunpeng@pdx.edu}

\author{Xueyan Liu}
\authornote{Corresponding author.}
\affiliation{
  \institution{Jilin University}
  \city{Changchun}
  \country{China}}
\email{xueyanliu@jlu.edu.cn}

\author{Minghao Yin}
\affiliation{%
  \institution{Northeast Normal University}
  \city{Changchun}
  \country{China}}
\email{ymh@nenu.edu.cn}







\renewcommand{\shortauthors}{Wenzhuo Song, et al.}

\begin{abstract}
Most session-based recommender systems (SBRSs) focus on extracting information from the observed items in the current session of a user to predict a next item, ignoring the causes outside the session (called outer-session causes, OSCs) that influence the user's selection of items. 
However, these causes widely exist in the real world, and few studies have investigated their role in SBRSs.
In this work, we analyze the causalities and correlations of the OSCs in SBRSs from the perspective of causal inference.
We find that the OSCs are essentially the confounders in SBRSs, which leads to spurious correlations in the data used to train SBRS models.
To address this problem, we propose a novel SBRS framework named COCO-SBRS (\textbf{CO}unterfactual \textbf{CO}llaborative \textbf{S}ession-\textbf{B}ased \textbf{R}ecommender \textbf{S}ystems) to learn the causality between OSCs and user-item interactions in SBRSs.
COCO-SBRS first adopts a self-supervised approach to pre-train a recommendation model by designing pseudo-labels of causes for each user's selection of the item in data to guide the training process.
Next, COCO-SBRS adopts counterfactual inference to recommend items based on the outputs of the pre-trained recommendation model considering the causalities to alleviate the data sparsity problem.
As a result, COCO-SBRS can learn the causalities in data, preventing the model from learning spurious correlations.
The experimental results of our extensive experiments conducted on three real-world datasets demonstrate the superiority of our proposed framework over ten representative SBRSs.
  
\end{abstract}



\maketitle

\section{Introduction}
\label{intro}
Existing studies on Recommender Systems (RSs) mainly focus on modeling and predicting all user-item interactions to learn users' static item preferences, ignoring the dynamic nature of user preferences in the real world~\cite{wang2022sequential}. 
To bridge this gap, \textit{session-based recommender systems (SBRSs)} have recently emerged and gained increasing attention \cite{wang2021survey,wang2020hierarchical}.
Most existing SBRS models formalize the recommendation of next user-item interactions (\textit{next item} for short) in sessions as a supervised learning task. For a current session of a user to be predicted (called \textit{target session}), an SBRS model takes the items of the observed interactions in this session (called the session's \textit{context}) as input and outputs a list of items as predicted next items \cite{wang2021survey}.
Most of these SBRSs imply a strong assumption that a user selects an item as the next item in a session only because the item correlates with the items in the same session.
However, they ignore the selection of the next items caused by factors other than items in the same session \cite{jannach2017session}.
For example, on an e-commerce site, a user may select a product, i.e., an item, because of his/her short-term intent or previous interest, a discount for this product, or a short-term item popularity trend. 
In either case, the item selected is not related to other items in the same session.

\begin{figure}[]
  \centering
  \includegraphics[width=\linewidth]{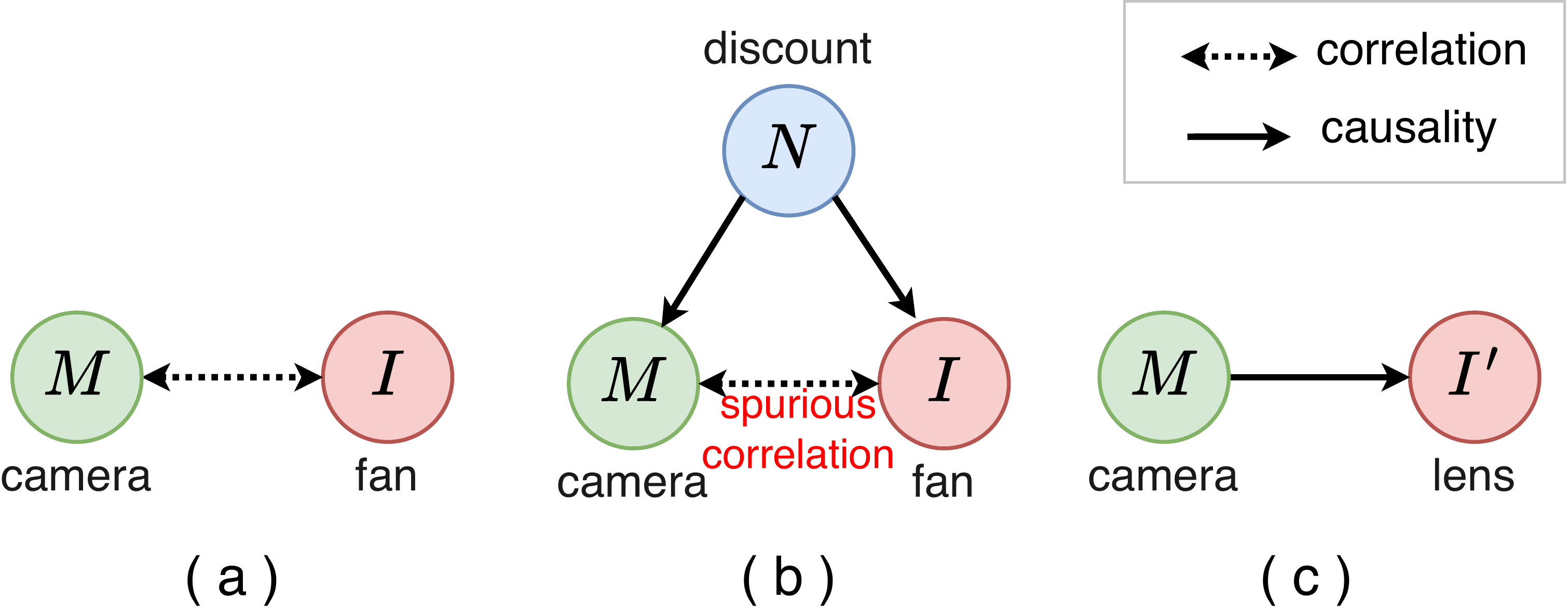}
  \caption{(a) Existing SBRSs make recommendations based on the co-occurrence-based correlation between camera and fan. (b) However, the correlation between camera and fan is spurious since the purchase of fan is not because of the purchase of camera. Instead, the co-purchase of camera and fan rely on the fact that they have discounted prices at the same time. (c) Actually, there are causalities between items in SBRS data. For example, a user buys a lens because he/she has bought a camera.}
  \Description[Causality and spurious correlation in SBRSs.]{Existing SBRSs make recommendations based on the co-occurrence-based correlations. However, the correlation may be spurious caused by the confounders in SBRSs. Actually, there are causalities in the data.}
  \label{fig:causal}
\end{figure}

In contrast, in this work, we suggest that it is not appropriate to train SBRSs without considering the influence of causality of the factors outside the session context. 
Let us consider that a user selects the next items in a session based on two types of causes, i.e., inner-session causes and outer-session causes. 
\textit{Inner-session causes} (\textit{ISCs} for short) refer to the selection of the next items caused by the context of the same session, which may be due to the co-occurrence pattern and sequential pattern between items in the same session.
In contrast, \textit{outer-session causes} (\textit{OSCs} for short) refer to the selection of the next item caused by factors outside the target session, such as long-term user preferences, product popularity trends and discounts for products on e-commerce sites.
To facilitate an insightful analysis, we first present a graph in Figure \ref{fig:causal} (a) showing the \textit{correlations} between variables considered in existing SBRS models.
For example, when camera and fan are on discount at the same time in a supermarket, many users buy them together because they are cheaper than usual.
Suppose camera and fan are ISCs $M$ and next item $I$ depicted in Figure \ref{fig:causal} (a), respectively.
The SBRS models learn the co-occurrence pattern $M\leftrightarrow I$ in the sessions.
If a user buys a camera, the models will recommend a fan to him/her even if the discount has ended, because there are a large number of sessions/transactions containing camera and fan together.
However, camera and fan are not related in terms of purposes. A user buys fan is not because he/she has bought camera, but because fan is on discount.
Therefore, this co-occurrence correlation may result in the mis-modeling problem in SBRSs.

To further explain this problem, we present two causal graphs to describe the \textit{causalities} in Figure \ref{fig:causal} (b) and (c). The causal graphs are directed acyclic graphs defined based on the structural causal model \cite{glymour2016causal}.
We use the causal graph in Figure \ref{fig:causal} (b) to represent the causalities in the above-mentioned example, where $N$ denotes the discount of a product, which is one of OSCs in SBRSs.
$N\rightarrow I$ and $N\rightarrow M$ denote a user buys a fan and a camera due to their discounted price respectively. Hence, the discounted price is the common cause of the purchase of camera and fan, and it is called a confounder\footnote{Confounders are variables that affect both the treatment variable, e.g., camera and the outcome variable, e.g., fan and lens.} in SBRSs \cite{pearl2018book,yao2021survey}. Clearly, the confounder results in a spurious correlation between camera and fan as depicted in Figure \ref{fig:causal} (b) since the purchase of fan is not because the purchase of camera even though they have been purchased successively in one session. A reliable SBRS should not make recommendations based on such spurious correlation since it cannot reflect the true casual relations between items.     
Actually, there are some other reliable causality relations between items in SBRS data. For example, as depicted in Figure \ref{fig:causal} (c), the purchase of a lens is often because of the prior purchase of a camera. Such kind of relations should be the basis for making reliable recommendations.  
From above examples, we can conclude that the OSCs can be the confounders in SBRSs, and the relationship between items in session contexts and the next item in the data is a mixture of the causality, e.g., camera and lens in Figure \ref{fig:causal} (c), and the spurious correlations, e.g., camera and fan in Figure \ref{fig:causal} (b).
Therefore, \textbf{(Problem 1)} \emph{without considering the causalities of the OSCs, the SBRSs will learn both the causality and the spurious correlations caused by the confounder and thus may generate incorrect recommendations}.

One possible way to solve this problem is to list all possible causes of the next item selection in sessions and then train a deep learning model in an end-to-end manner to learn the complex relationship between the next item and the causes.
However, the \textit{true cause}, i.e., which of ISCs and OSCs is the reason why a user selects a specific item, is a latent variable in the model, and they are not provided in the dataset.
Thus, \textbf{(Problem 2)} \emph{it is difficult for deep learning models to identify the true causes of the item selections and learn the causalities shown in Figure \ref{fig:causal}} (b) and (c).

In this paper, we \ul{address Problem 1} by proposing a novel SBRS framework considering both OSCs and ISCs for the next item recommendation in sessions. The proposed model is named \emph{\textbf{CO}unterfactual \textbf{CO}llaborative \textbf{S}ession-\textbf{B}ased \textbf{R}ecommender \textbf{S}ystem (COCO-SBRS, or COCO for short)}. 
COCO is inspired by the ideas of counterfactuals \cite{yao2021survey} and collaborative filtering for sessions \cite{sknn,stan}.
Specifically, COCO first pre-trains a recommendation model to learn the causalities among ISCs, OSCs and user-item interactions in SBRSs, and then predicts the next item for a session based on some neighbor sessions with the same ISCs and OSCs as this session and the recommendation model used to simulate the user's selection in the neighbor sessions.
To \ul{address Problem 2}, in the pre-training phase, we adopt a self-supervised approach to train the recommendation model by designing pseudo-labels of causes for each user-item interaction to guide the training of the model.
To alleviate the problem of the lack of sessions with required ISCs and OSCs in the prediction phase, we adopt counterfactual inference to generate sessions using required ISCs and OSCs. This simulates users' decision-making process with the pre-trained recommendation model to recommend items.
In summary, the main contributions of this work are:
\begin{enumerate}
    \item We propose an SBRS framework named counterfactual collaborative session-based recommender system (COCO-SBRS or COCO) for effective next item recommendations in sessions.
    To the best of our knowledge, this is the first work in the literature to address the problem of spurious correlations caused by the confounder in SBRSs.
    \item We are the first to formulate the next item recommendation in sessions in the framework of counterfactual computing and collaborative filtering. Specifically, we first develop a self-supervised approach to pre-train a recommendation model to learn causalities in SBRSs. Then, we recommend items for a given session using the model, taking other sessions with the same ISCs and OSCs as this session as input.
    \item To evaluate the effectiveness of COCO, we conduct extensive experiments on three real-world datasets from various domains. We compare COCO with the representative and state-of-the-art SBRSs, and the experimental results show COCO's superiority over baseline SBRSs.
\end{enumerate}

\section{Related Work}
\subsection{Session-based Recommender Systems}
In this section, we introduce two groups of SBRSs: (1) SBRSs that consider inner-session causes only, and (2) SBRSs that consider both inner-session and outer-session causes.

\textbf{SBRSs that consider inner-session causes only} make recommendations based on the session context.
These algorithms aim to learn users' short-term preferences reflected by the items in the session context and the complex relationships between items in sessions.
According to the employed technology, these SBRSs can be classified into conventional approaches, latent representation approaches, and deep neural network approaches \cite{wang2021survey}.
KNN-based SBRSs and Markov chain-based SBRSs are the most popular conventional approaches.
KNN-based SBRSs such as SKNN recommend items in sessions similar to the target session \cite{sknn,stan}.
Markov chain-based SBRSs such as FPMC make recommendations by modeling the transition patterns of items in sessions \cite{fpmc,le2016modeling}.
The latent representation SBRSs utilize the technique of latent factors models or matrix factorization to make recommendations \cite{liu2013personalized,cheng2013you,SONG201842,SONG2021329}.
In recent years, deep learning-based SBRSs are becoming popular, and researchers have developed SBRSs based on various deep learning techniques.
Recurrent neural network (RNN) based SBRSs model the sequential pattern of items in sessions \cite{gru4rec,hidasi2018recurrent,wu2017recurrent}. However, they are based on a rigid assumption that adjacent items in each session are sequentially dependent.
Attention-based methods relax this assumption by emphasizing those more informative items in sessions to reduce the interference of uninformative ones \cite{chen2019dynamic,ying2018sequential,wang2022veracity}.
To model the high-order transition among items, the graph neural network (GNN) based SBRSs first represent sessions with graphs and then employ GNN models to make recommendations \cite{srgnn,qiu2019rethinking}.
Deep learning-based SBRSs can model the complex patterns of items in sessions and therefore achieve better performance over other approaches in many recent studies.
However, the above-mentioned SBRSs can only predict the next item based on the limited information in the session context and ignore the global information and factors outside the session influence users' selection of items.

\textbf{SBRSs that consider both inner-session and outer-session causes} aim to extract and fuse information from both the target session and other sessions to make recommendations. 
The most common information from other sessions considered in existing works includes global information such as item dependency and long-term item preferences of users.
Global information is essential for SBRSs because the lengths of sessions in the real world tend to be very short and contain limited information, and the global information can be used as a supplement to the session context \cite{qiu2020exploiting,wang2022exploiting}.
User preferences are outer-session information that has great potential to improve the performance of recommendation algorithms because, in many real-world scenarios, it is easy to obtain user behavioral data reflecting user preferences that belong to the same user of the target session \cite{insert,li2018learning,sun2019can}.
Unlike the above algorithms, this work aims to eliminate the spurious correlations caused by the confounder in SBRSs introduced in Section \ref{intro}.

\subsection{Counterfactuals in Recommender Systems}

Counterfactual computing has been introduced into the research of recommendation systems in the recent years. Compared to the rapid development in other machine learning fields, the studies on counterfactual-based recommender systems (CFBRSs for short) are still limited.
According to the problem to be addressed, existing CFBRSs can be classified into algorithms for data bias, algorithms for data missing, and algorithms for other tasks \cite{survey2022causalrs,wang2022trustworthy}.
Among the algorithms for data bias, researchers have proposed the MACR algorithm for popularity bias, and the CR algorithm for clickbait bias \cite{wei2021model,wang2021clicks,mu2022alleviating,chen2021autodebias,liu2021mitigating}.
The data missing problem stems from the large number of inactive users and items in RSs. Existing works use counterfactual-based data generation to alleviate the problem of missing data in RSs \cite{wang2021counterfactual,zhang2021causerec,xiong2021counterfactual}.
Other problems studied in CFBRSs include explainability, diversity and fairness in recommender systems \cite{tran2021recommending,ge2022explainable,tan2021counterfactual,wang2022user,zhang2021counterfactual}.
Different from the existing CFBRSs, we study the counterfactual-based model for learning causality in session-based recommender systems.

\section{Problem Statement}

In this work, we use $U=\{u_1,...,u_{|U|}\}$ to denote the set of $|U|$ users and $V=\{v_1,...,v_{|V|}\}$ to denote the set of $|V|$ items in the dataset.
A user $u\in U$ has a sequence of user-item interactions, which can be represented using a list of items corresponding to each interaction.
In SBRSs, the interactions of each user form a series of sessions, where each session is a list of the user's interactions, i.e., item list $s=\{v_1^s,...,v_t^s\}$ in a short period of time. The subscript of each item in $s$ denotes its order in $s$ w.r.t its occurring time.

Given a target session $s$ to be predicted, the goal of an SBRS is to predict the next interactions in $s$, i.e., the next item $v_{t+1}^s$, based on known information such as user preferences and session context.
In this work, we consider two causes for a user to select the next item in sessions, i.e., ISCs and OSCs.
Specifically, we formulate the SBRS model as a probabilistic classifier $p(v|M, N)$ over a set of items in the dataset conditioning on the ISCs variable $M$ and the OSCs variable $N$.
The prediction result of $p(v|M, N)$ represents the probability distribution of the item which user will select as the next item for $s$.
Finally, items with top-K probabilities are selected as the recommendation result.



\section{Counterfactual Collaborative Session-based Recommender System}
\begin{figure*}[]
  \centering
  \includegraphics[width=0.9\linewidth]{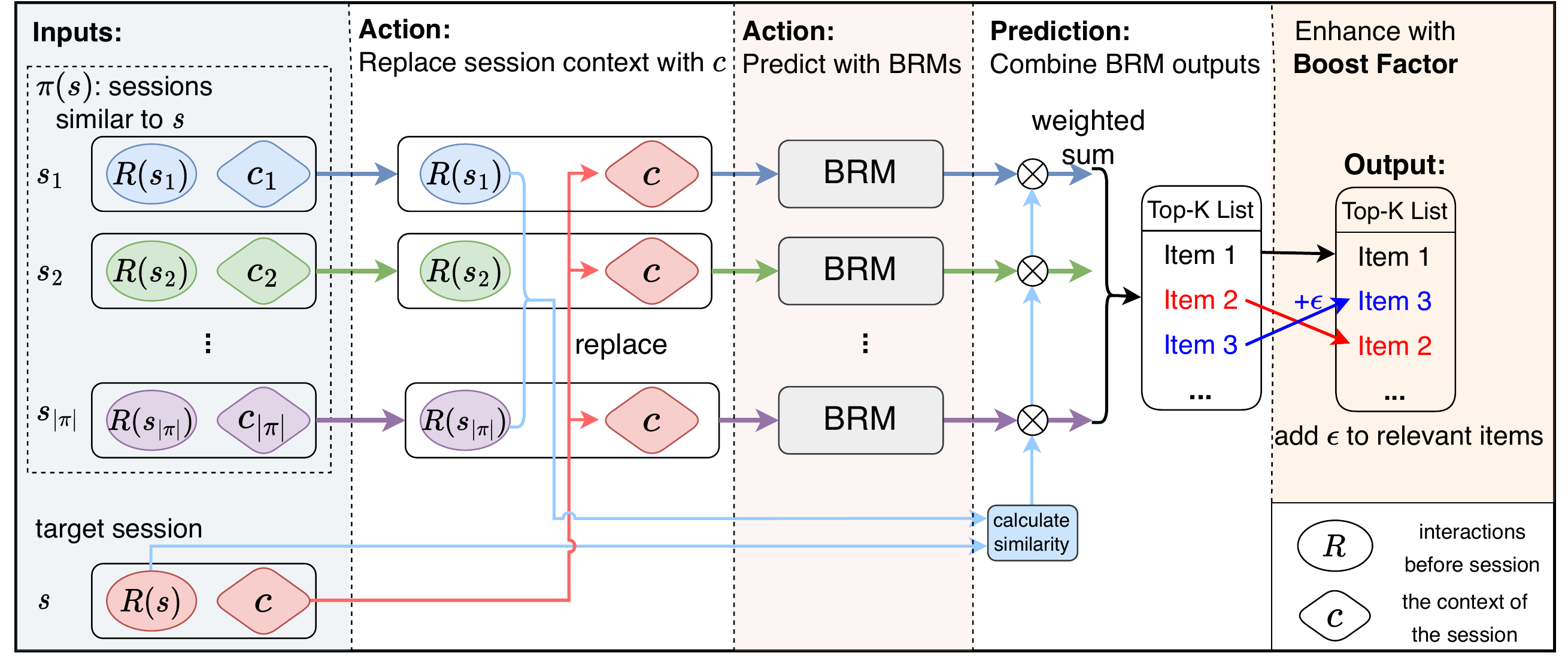}
  \caption{The steps of Action and Prediction of COCO-SBRS. The BRMs are pre-trained with the sessions in the training set in the Abduction step, which is not shown in this figure.}
  \Description[The steps of action and prediction of COCO-SBRS.]{In action step, COCO replaces the ISCs of each sessions, and used them as the input to the BRMs. In prediction step, COCO makes recommendations by combing the outputs of all the BRMs.}
  \label{fig:framework}
\end{figure*}

In this section, we introduce the proposed SBRS method named \textbf{CO}unterfactual \textbf{CO}llaborative \textbf{S}ession-\textbf{B}ased \textbf{R}ecommender \textbf{S}ystem (COCO-SBRS, or COCO for short).
COCO is inspired by counterfactuals \cite{yao2021survey,glymour2016causal} and collaborative filtering for sessions \cite{sknn,insert}.
COCO first pre-trains a recommendation model to learn the causalities among ISCs, OSCs and user-item interactions in SBRSs, and then predicts the next item for a session based on some neighbor sessions with the same ISCs and OSCs as this session and the recommendation model used to simulate the user's selection in the neighbor sessions.
Specifically, COCO has the following three steps:
\textbf{(i) Abduction}: 
In this step, we design a generative process to describe the users' decision-making process in SBRSs, and then implement the generative process with an attention-based neural network.
Next, we adopt a self-supervised approach to pre-train the neural network using the sessions in the training set to determine its optimal parameters.
After pre-training the BRM, the framework makes recommendations with the steps of action and prediction, the process of which is shown in Figure \ref{fig:framework}.
\textbf{(ii) Action}: For a target session $s$ of user $u$, we first sample a session $s'$ of user $u'$ from the dataset, then replace its ISCs, i.e., $ISC(s')$ with the ISCs of the target session, i.e., $ISC(s)$.
Next, we use the modified session as the input to the model trained in the abduction step to simulate the decision-making process of $u'$ to answer a counterfactual question \cite{yang2021top}:
``\textit{Given a target session $s$ of $u$, which item will be selected as the next item by another user $u'$ given the ISCs of $s$?}''
We sample a few sessions of users similar to $u$, and then perform the action step for each of them.
\textbf{(iii) Prediction}:
In this step, we make recommendations by combining the outputs of all the models, i.e., the answers to the counterfactual question, and the users' similarities.

\subsection{Abduction: Base Recommendation Model and its Self-Supervised based Training}


\begin{figure}[]
  \centering
  \includegraphics[width=1\linewidth]{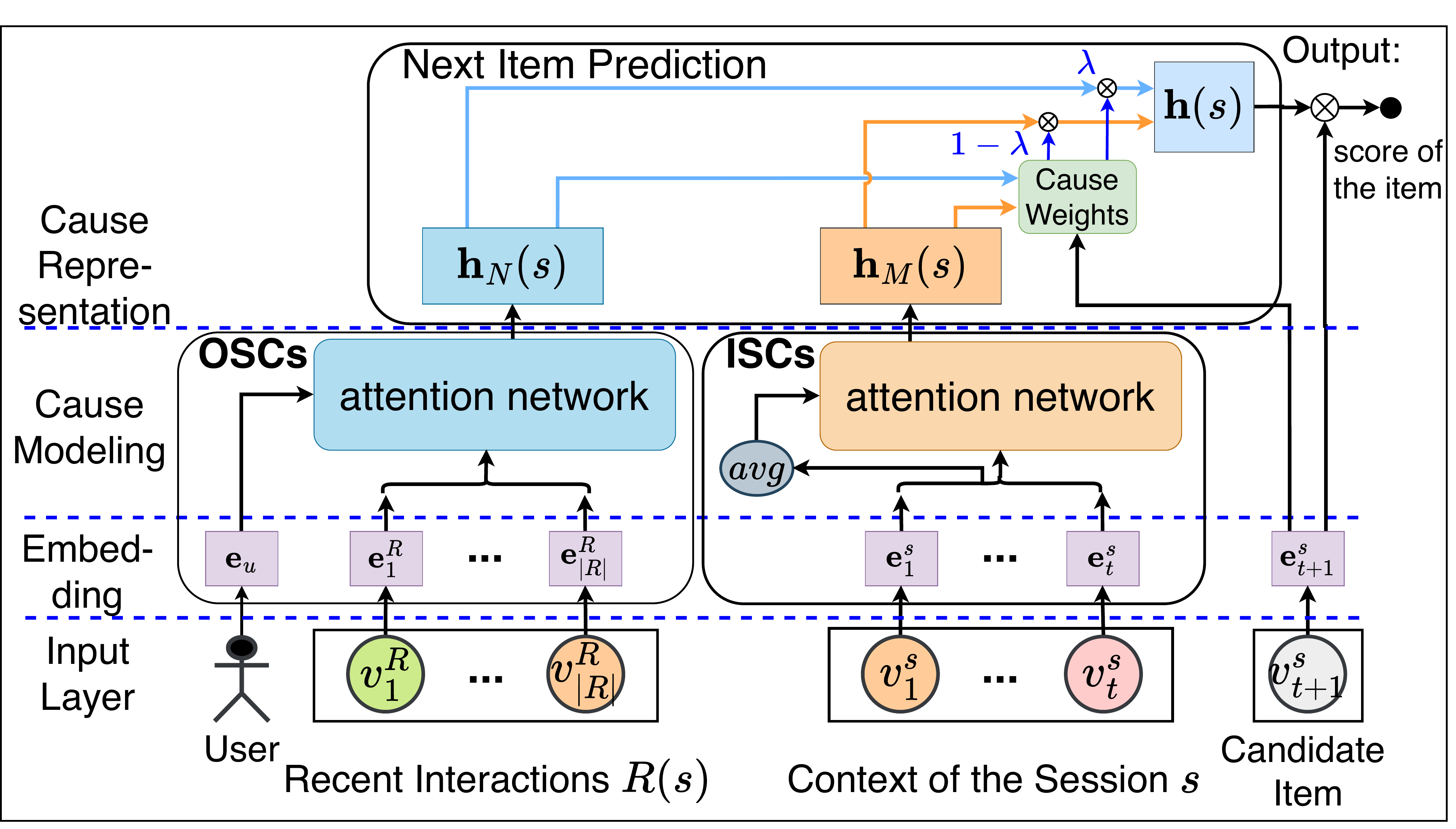}
  \caption{Base Recommendation Model (BRM).}
  \Description[Base recommendation model (BRM).]{The BRM is an attention-based recommendation model which contains a recent interactions encoder, a session context encoder, and an attention based embedding prediction module.}
  \label{fig:brm}
\end{figure}

\textbf{Generative Process for SBRSs:}
Based on the causal graphs in Figure \ref{fig:causal} (b) and (c), we present a generative process to describe a user's decision-making process given specific ISCs and OSCs:

Given a target session $s$ of user $u$:
\begin{enumerate}
    \item generate $OSC(s)$ according to the user $u$ of $s$;
    \item generate $ISC(s)$ according to the items in $s$'s context $c$;
    \item sample a cause $C(u,c)$ from $OSC(s)$ and $ISC(s)$ according to $u$ and $c$ for an item $v$ to predict;
    \item generates the score of the item based on $C(u,c)$.
\end{enumerate}
Next, we implement the generative process for SBRSs based on attention neural networks.
We call the model the base recommendation model (BRM). BRM's architecture is depicted in Figure \ref{fig:brm}.

\textbf{Modeling OSCs in BRM:}
We consider two OSCs in SBRSs: static preference and dynamic preference \cite{wang2021survey}. 
(1) Static preference is a long-term user preference that does not change with time. 
In this work, we use a fixed-length vector, i.e., user embedding $\textbf{e}_u\in \mathbb{R}^d$, to represent the static preference of a user $u$. 
(2) Dynamic preference is the short-term user preference that changes with time. 
We use an attention-based neural network to learn the embedding of a user's dynamic preference.
Specifically, given a user $u$'s session $s$, the embedding of dynamic preference $\textbf{h}_N(s)$ for $s$ is calculated by:
\begin{equation}
    \textbf{h}_N(s)=attention(\textbf{e}_u,\textbf{E}(R(s)),\textbf{E}(R(s))),
\end{equation}
where $R(s)=\{v_1^R,...,v_{|R|}^R\}$ is a set containing the most recent interactions of $u$ before $s$, $|R(s)|$ is a hyper-parameter and is set as 10 for all sessions in our experiments,
$\textbf{E}(R(s))\in\mathbb{R}^{|R(s)|\times d}$ denotes the matrix containing all the embeddings of items in $R(s)$. Given a query vector $\textbf{q}'\in \mathbb{R}^d$, a matrix of $\kappa$ key vectors $\textbf{K}'\in\mathbb{R}^{\kappa\times d}$ and a matrix of $\kappa$ value vectors $\textbf{V}'\in\mathbb{R}^{\kappa\times d}$, $attention(\textbf{q}', \textbf{K}', \textbf{V}')$ is an attention network defined as:
\begin{equation}
    attention(\textbf{q}',\textbf{K}',\textbf{V}') = \sum_i \alpha(\textbf{q}',\textbf{K}'_i)\times \textbf{V}'_i,
\end{equation}
where $\alpha(\textbf{q}',\textbf{K}'_i)$ is a vector containing the attention scores between $\textbf{q}'$ and each vector in $\textbf{K}'$ calculated by:
\begin{equation}
    \alpha(\textbf{q}',\textbf{K}'_i) = softmax_i( \textbf{K}'_i\cdot\textbf{q}'^{T}/\sqrt{d}).
\end{equation}

Finally, we use $\textbf{h}_N(s)$ as the embedding of the OSCs for $s$.

\textbf{Modeling ISCs in BRM:}
Given $s$'s context $c$, the representation of ISCs, i.e., $\textbf{h}_M(s)$ for $s$ is calculated by:
\begin{equation}
    \textbf{h}_M(s)=attention(avg(\textbf{E}(c)),\textbf{E}(c),\textbf{E}(c)),
\end{equation}
where $\textbf{E}(c)\in \mathbb{R}^{|c|\times d}$ denotes a matrix containing the embeddings of all items in $c$, $avg(\textbf{E}(c))=\frac{1}{|c|}\sum_{v\in c}\textbf{e}_v$ is the mean vector of the item embeddings in $c$.

\textbf{Next-Item Prediction in BRM:}
In the generation process, the user selects one of the OSCs and ISCs as the cause for the next item selection in $s$ according to the user and the session.
We implement this process with an attention-based neural network to learn a soft weight of the two causes, i.e., $\lambda\in[0,1]$:
\begin{equation}
    \label{eq:causalattn}
    \lambda=sigmoid(\textbf{W}^T(\textbf{h}_M(s)||\textbf{h}_N(s)||\textbf{e}_v)+b),
\end{equation}
where $sigmoid(z)=1/(1+e^{-z})$, $||$ denotes the concatenate operation, $v\in V$ is the item to be predicted, and $\textbf{W}\in \mathbb{R}^d$ and $b\in \mathbb{R}$ are parameters to learn. A large $\lambda$ means that the OSCs have more influence than the ISCs on the next item selection, and a small $\lambda$ means the ISCs have more influence than the OSCs for $s$.

Next, we incorporate the two causes by:
\begin{equation}
    \textbf{h}(s)=\lambda \textbf{h}_N(s) + (1-\lambda) \textbf{h}_M(s),
\end{equation}
and predict the next item for $s$ by:
\begin{equation}
    \label{eq:brm_pred}
    p(v|M=ISC(s),N=OSC(s))=softmax(\textbf{e}(\Omega)\textbf{h}(s)^T),
\end{equation}
where $\Omega$ is the set of items to be predicted, and $\textbf{e}(\Omega)\in \mathbb{R}^{|\Omega|\times d}$ is a matrix containing the embeddings of all items in $\Omega$.

\textbf{Training BRM with Cross Entropy Loss:}
By regarding the next item prediction in sessions as a multi-class classification task, we can train the model using the Cross Entropy Loss:
\begin{equation}
    \label{eq:cel}
    \mathscr{l}_1=-\frac{1}{|S_b|}\sum_{s\in S_b}[\log p(v=v^+)+\sum_{v'\in V_b,v'\neq v^+}\log(1-p(v=v'))],
\end{equation}
where $p(v)$ is short for $p(v|M=ISC(s),N=OSC(s))$,$v^+$ is the true next item of $s$, $S_b$ denotes the set of all sessions in the training batch, $V_b$ is the set of all items in the training batch.


\textbf{Improve BRM with Self-Supervised Learning:}
In the proposed generation process, an item is selected as the next item due to one of the ISCs and OSCs, which we call the \textit{true cause} for the corresponding user-item interaction.
However, the true cause is a latent variable in the model since they are not provided in the dataset.
To address this problem, in BRM, we use a linear model, i.e., Equation (\ref{eq:causalattn}), to learn the relationship between the latent variable $\lambda$ and the observational variables $M$ and $N$ so that the model gets the ability to identify the true cause corresponding to each interaction.

To improve the accuracy of true cause identification, we propose a self-supervised learning approach to guide the training of the parameters in Equation (\ref{eq:causalattn}).
Our main idea is to construct a pseudo-label for the soft weights of two causes, i.e., $\lambda$. Specifically, we assume that: given each item of each session, (1) if an item $v$ appears in the interaction history of $u$, the user is more likely to select the item due to the OSCs, i.e., $\lambda=1$; (2) if $v$ appears in the context of the target session, the user is more likely to select the item due to the ISCs, i.e., $(1-\lambda)=1$. 
Based on these two assumptions, we obtain the definitions of $\lambda$'s pseudo-labels:
\begin{equation}
\label{eq:1}
\begin{split}
\begin{cases}
y_N(s,v) = \mathit{1}[v \in V_u],\\
y_M(s,v) = \mathit{1}[v \in c],
\end{cases}
\end{split}
\end{equation}
where $V_u$ is the set of items the user of $s$ has interacted before, and $\mathit{1}[condition]$ is defined as:
\begin{equation}
\label{eq1}
\begin{split}
\mathit{1}[condition] =
\begin{cases}
1,& \text{if $condition$ is True,}\\
0,& \text{if $condition$ is False.}
\end{cases}
\end{split}
\end{equation}

Next, we consider the true cause prediction problem as a binary classification problem and present a self-supervised loss based on the Binary Cross Entropy Loss:
\begin{equation}
    \label{eq:l2}
    \mathscr{l}_2=\frac{1}{|S_b|}\sum_{s\in S_b}BCE(\lambda,y_N(s,v^+))+BCE((1-\lambda),y_M(s,v^+)),
\end{equation}
where $v^+$ is the true next item of $s$, and given a prediction score $x$ and the corresponding label $y$, Binary Cross Entropy Loss $BCE(x,y)$ is defined as:
\begin{equation}
    BCE(x,y)=y\cdot \log x + (1-y)\cdot \log (1-x).
\end{equation}

Training BRM according to Equation (\ref{eq:l2}) encourages the model to generate soft weights of the OSCs and ISCs that are consistent with pseudo-labels, improving the model's ability to untangle the ISCs and OSCs, and identify the true causes of the interactions. 

Finally, the loss function for the training of BRM is defined as a trade-off of two loss functions:
\begin{equation}
    \mathscr{l}=\mathscr{l}_1+\beta*\mathscr{l}_2,
\end{equation}
where $\beta\in [0,+\infty)$ is the trade-off hyper-parameter.

\subsection{Action and Prediction: Counterfactual and Collaborative Next-Item Recommendation}

The key idea of collaborative filtering for sessions is predicting items in other sessions similar to the target session \cite{sknn,stan}.
However, when calculating session similarities, these methods only consider the items in the session context, and ignore outer-session causes such as static user preference.
In addition, these methods assume that every session in the dataset has similar sessions, which may not be true when the dataset is sparse.
As a result, they may treat sessions of users who have a completely different preference as similar sessions, and thus make incorrect recommendations.

To address the above problems, given a target session $s$ of a user $u$, we use the BRM model to simulate users' decision-making process for the counterfactual question: ``which next item will be selected by another user $u'$, who has similar preferences to $u$ (i.e., $OSC(s)$ in BRM) in the same session context (i.e., $ISC(s)$ in BRM)?''
In this way, we can ensure that the next item for each similar session is generated when the user has similar ISCs and OSCs to the target session.
Specifically, the process contains two steps:

\textbf{Action:} Given a target session $s$ of a user $u$, the action step aims to find a user $u'$ who has a similar preference to $u$ and then compute the probability distribution of the next item under the same ISCs of $s$. 
Specifically, we first find a session $s'$ with similar OSCs to $s$ by calculating the similarity of the recent interaction sets\footnote{The most recent interaction set of a session, i.e., $R(s)$, can reflect the dynamic user preferences when the session occurs, which is the main part of the OSCs for $s$.}:
\begin{equation}
    \label{eq:sess_sim}
    sim(s,s')=\frac{|R(s)\cap R(s')|}{|R(s)\cup R(s')|},
\end{equation}
where $s'\neq s$ is a session sampled from the dataset. 
A large $sim(s,s')$ means that $u'$ has similar OSCs to $u$ when the session $s'$ occurred.

However, a larger value of $sim(s,s')$ does not mean that the ISCs of $s'$ are similar to the ISCs of $s$, and direct use of BRM to predict for $s$ based on $s'$ will produce erroneous results.
We replace the $ISC(s')$ with $ISC(s)$ as an action on the BRM's prediction in Equation (\ref{eq:brm_pred}) to address this problem:
\begin{equation}
   p(v|M=ISC(s),N=OSC(s')).
\end{equation}
In this way, the BRM can predict the next item as it simulates $u'$ in the context of session $s$, which does not exist in the dataset.

\textbf{Prediction:}
In the action step, we only consider one similar user.
Based on the idea of collaborative filtering, in the prediction step, we consider that increasing the number of similar users could improve the recommendation performance of our model.

Specifically, we first select the most similar sessions of $s$ from the dataset to form a session set $\pi(s)$ according to the session similarities of Equation (\ref{eq:sess_sim}). Next, for each user $u_i$ of session $s_i\in \pi(s)$, we replace the $ISC(s_i)$ in its corresponding BRM with $ISC(s)$ while keeping its $OSC(s_i)$ unchanged, and then we weighted sum the results of all BRMs of each session in $\pi(s)$ to obtain the result:
\begin{equation}
\begin{split}
    \label{eq:coco}
    p_{coco}(v&|M=ISC(s),N=OSC(s))=\\
    &\frac{1}{C}\sum_{s_i\in \pi(s)} sim(s,s_i) \times p(v|M=ISC(s),N=OSC(s_i)),
\end{split}
\end{equation}
where $C$ is the normalization factor.

\subsection{Enhancement with Boost Factor}
After obtaining the item distribution $p_{coco}(v|M,N)$ for the next item recommendation, we find that emphasizing the recently seen items in the target session $s$ can further improve the performance of our model. Specifically, for each item $v_i\in (R(s)\cup c)$, we first modify the weights of recently seen items by adding a boost factor $\epsilon>0$ and then obtain the final result of COCO after enhancement with the boost factor
:

\begin{equation}
\begin{split}
    p'_{coco}(v&|M=ISC(s),N=OSC(s))=\\
    &\frac{1}{C'}(p_{coco}(v|M,N)+\epsilon*\textbf{1}[v\in (R(s)\cup c)]),
\end{split}
\end{equation}
where $\textbf{1}[condition]$ is a vector containing the results of Equation (\ref{eq1}) for all $v\in V$, and $C'$ is the normalization factor.

\begin{table}[t]
\centering
\caption{The detailed statistical information of the three datasets used in the experiments.}
\label{table:dataset}
\begin{tabular}{lrrr}
\hline
                            & Last.fm & Delicious & Reddit    \\ \hline
\#sessions                  & 5915    & 45,772    & 1,122,150 \\
\#interactions              & 38,367  & 249,919   & 2,874,671 \\
\#users                     & 1,101   & 1,752     & 19,878    \\
\#items                     & 711     & 5,047     & 13,742    \\
\#interactions per user    & 34.85   & 142.65    & 144.62    \\
\#interactions per session & 6.49    & 5.46      & 2.56      \\
\#sessions per user         & 5.37    & 26.13     & 56.45     \\ \hline
\end{tabular}
\end{table}

\section{Experiments}
\begin{table*}[]
\centering
\caption{Recommendation performance of all compared methods on three datasets. R@5 and R@20 are short for Recall@5 and Recall@20. N@5 and N@20 are short for NDCG@5 and NDCG@20. We adopt 5-fold cross-validation and report the average value for each metric. For all evaluation metrics, higher numbers represent better method performance. The bold and underlined numbers under each metric represent the best and the second-best performing method, respectively. * means the improvement is significant at $p<0.05$.}
\label{table:exp1}
\begin{tabular}{|l|cccc|cccc|cccc|}
\hline
        & \multicolumn{4}{c|}{\textbf{Last.fm}}                                       & \multicolumn{4}{c|}{\textbf{Delicious}}                                     & \multicolumn{4}{c|}{\textbf{Reddit}}                                        \\ \hline
        & \textbf{R@5} & \textbf{N@5} & \textbf{R@20} & \textbf{N@20} & \textbf{R@5} & \textbf{N@5} & \textbf{R@20} & \textbf{N@20} & \textbf{R@5} & \textbf{N@5} & \textbf{R@20} & \textbf{N@20} \\ \hline
SKNN    & 0.235             & 0.116           & 0.536              & 0.202            & 0.111             & 0.055           & 0.293              & 0.107             & 0.202            & 0.107           & 0.383              & 0.159          \\
GRU4Rec & 0.331             & 0.238           & 0.542              & 0.298            & 0.234             & 0.172          & 0.390              & 0.216            & 0.225             & 0.163          & 0.397              & 0.212            \\
STAMP   & 0.269             & 0.190           & 0.500              & 0.256            & 0.150             & 0.104         & 0.294             & 0.105            & 0.192             & 0.133           & 0.323              & 0.170            \\
CSRM    & 0.342             & 0.250           & 0.562              & 0.312            & 0.197             & 0.144           & 0.346              & 0.186            & 0.200             & 0.150           & 0.366              & 0.197            \\
SR-GNN   & 0.265             & 0.186           & 0.477              & 0.246            & 0.205             & 0.149           & 0.354              & 0.191            & 0.251             & 0.186           & 0.422              & 0.235            \\ \hline
HGRU    & 0.340             & 0.242           & 0.576              & 0.309            & 0.218             & 0.160           & 0.377              & 0.205            & 0.337             & 0.257           & 0.518              & 0.309            \\
II-RNN   & 0.359             & \underline{0.259}           & 0.586              & \underline{0.323}            & 0.257             & 0.189           & 0.424              & 0.236            & 0.365             & \underline{0.277}           & 0.560              & 0.333            \\
SASRec & 0.346 & 0.206 & \underline{0.651} & 0.201 & 0.193 & 0.130 & 0.385 & 0.184 & 0.310 & 0.209 & 0.537 & 0.274 \\
BERT4Rec & 0.304 & 0.182 & 0.622 & 0.273 & 0.207 & 0.101 & 0.369 & 0.186 & \underline{0.409} & 0.271 & \underline{0.623} & \underline{0.338}\\
INSERT  & \underline{0.364}             & 0.258  & 0.589             & \underline{0.323}            & \underline{0.264}             & \underline{0.196}  & \underline{0.436}             & \underline{0.245}   & 0.391             & \textbf{0.301}  & 0.561              & \textbf{0.350}   \\ \hline
COCO-SBRS       & \textbf{0.504*}    & \textbf{0.289*}           & \textbf{0.793*}     & \textbf{0.374*}   & \textbf{0.359*}    & \textbf{0.215*}           & \textbf{0.520*}     & \textbf{0.263*}            & \textbf{0.412*}    & 0.248           & \textbf{0.699*}     & 0.331            \\ 

\hline

\end{tabular}
\end{table*}
\subsection{Experimental Settings}
\textbf{Baseline Algorithms:} 
In the comparison experiments, we select (1) five SBRSs that consider inner-session causes only (SKNN, GRU4Rec, STAMP, CSRM and SRGNN) and (2) five SBRSs that consider both inner-session and outer-session causes (HGRU, II-RNN, SASRec, BERT4Rec and INSERT) as baseline algorithms.
These baselines are based on various techniques, including collaborative filtering, RNN, attention neural networks, memory neural networks and graph neural networks.
\begin{itemize}
    \item \textbf{SKNN}: A collaborative filtering SBRS built on the idea of recommending the items in neighbor sessions which are similar to the target session \cite{sknn}.
    \item \textbf{GRU4Rec}: GRU4Rec employs an RNN to extract dynamic user preference in the session context for the next item recommendation \cite{gru4rec}.
    \item \textbf{STAMP}: A memory neural network-based SBRS uses attention to capture users' short-term preferences for the next item recommendations in sessions \cite{stamp}.
    \item \textbf{CSRM}: A memory neural network-based SBRS uses attention networks to learn the similarities between sessions and make recommendations based on the information extracted from neighbor sessions and their similarities \cite{csrm}.
    \item \textbf{SR-GNN}: A graph neural network-based SBRS first represents the sessions with graphs and then employs a graph neural network to recommend items based on the item transition patterns extracted from the graph \cite{srgnn}.
\end{itemize}

The following five SBRSs consider both inter-session causes, i.e., the information from the session context, and outer-session causes, e.g., users preference:

\begin{itemize}
    \item \textbf{HGRU}: HGRU employs a hierarchical RNN where one RNN models the sequential patterns in the session context, and the other RNN learns a user's preference across his sessions \cite{hgru}.
    \item \textbf{II-RNN}: II-RNN utilizes the information extracted from the most recent session to complement and initialize the RNN modeling the target session \cite{iirnn}.
    \item \textbf{SASRec}: SASRec is a self-attention based sequential RS designed to model users' interaction sequences. In this work, we concatenate all interactions of each user to form the user sequences \cite{kang2018self}.
    \item \textbf{BERT4Rec}: BERT4Rec is a deep bidirectional self-attention based sequential RS model. BERT4Rec adopts the Cloze objective and predicts an item based on its context and the user's historical interactions  \cite{sun2019bert4rec}.
    \item \textbf{INSERT}: INSERT is the state-of-the-art SBRS considering both user preference and item patterns in sessions. It is designed for next item recommendations in short sessions based on few-shot learning and meta-learning \cite{insert}.
\end{itemize}

\textbf{Datasets:}
We conduct the experiments on the following three publicly available real-world datasets used in previous SBRS works:
\begin{itemize}
    \item \textbf{Last.fm}\footnote{Last.fm and Delicious are from https://grouplens.org/datasets/hetrec-2011/.} used in \cite{latifi2021session} contains the logs of users' music listening behaviors in Last.fm online music service \cite{dataset}.
    \item \textbf{Delicious} is a dataset used in \cite{song2019session} which contains the user tagging records in a social network-based bookmarking system named Delicious.
    \item \textbf{Reddit}\footnote{https://www.kaggle.com/datasets/colemaclean/subreddit-interactions} used in II-RNN \cite{iirnn} records the user's history of visiting the subreddits, i.e., discussion topics, in Reddit.
\end{itemize}

\textbf{Data Preparation:} We prepare the datasets based on the process in previous SBRS works \cite{latifi2021session,iirnn,insert}.
For each dataset, we remove inactive users and items with the number of interactions less than 10. 
Then, we put two successive items in a user's interaction history into one session if the interval between them is less than 6 hours. In contrast, if the interval is greater than 6 hours, they are put into two different sessions.
Next, we remove the sessions containing only one item and the sessions with more than 20 items. 
The reason for not considering sessions with one item is that these sessions cannot have the context together with a item to be predicted \cite{insert}.
Removing long sessions is a common practice in SBRSs \cite{latifi2021session,insert}, because there are few long sessions, so removing them will not affect the results of SBRSs, but keeping them will greatly increase the running time of many baseline algorithms.
After the pre-processing, we show the basic information of the three datasets in Table \ref{table:dataset}.


\textbf{Evaluation Protocol and Metrics:}
We use 5-fold cross-validation to obtain reliable experimental results. Specifically, we randomly divide all sessions in the dataset into five equal parts. We select one of the five parts as the test set and the remaining as the training set for each experiment. The experiments for each dataset will be conducted five times so that the test sets cover all the data, and the average experimental results of the five parts are reported.
Note that in each fold, we ensure that the training set contains all users and all items to avoid abnormal results.
Besides, we randomly select half of the sessions in the test set to form the validation set.

We use commonly used ranking measures in previous works, i.e., recall and NDCG, to evaluate the performance of each SBRS \cite{insert,sun2019bert4rec,kang2018self}.
Specifically, for each test session, we iteratively select each item as the target item and the items before this item is the session's context.
For each algorithm, we sort all items based on their recommendation scores output by the model and select the top-K scored items to calculate recall@K and NDCG@K.

To obtain the best performance for each algorithm, we first initialize its hyper-parameters using the settings given in the paper for each algorithm and then fine-tune the most important hyper-parameters based on the performance of each algorithm on the validation set.
We report the performance of the next item recommendations in short sessions since some SBRSs will perform better (see Appendix \ref{app_a} for details).
The main parameters of the baselines are set as follows: In KNN, the number of similar sessions is 500; In CSRM, the number of memory slots is 256; In II-RNN, the dimension of embeddings is 50; In INSERT, the dimension of the hidden state is 50, and the number of similar sessions is set to 10; 
In both SASRec and BERT4Rec, we set the number of attention layers $L=2$ and the number of attention heads $h=2$.
In the proposed COCO-SBRS, early stopping w.r.t. the R@20 on validation set is used with 50 as the maximum number of epochs. The learning rate is 0.01, the number of other sessions, i.e., $|\pi|$, is 10, and the trade-off parameter $\beta$ is set to 1 for all experiments. 
We implement COCO-SBRS using PyTorch and GPU to accelerate the model's training.
The embeddings of users and items are randomly initialized with the commonly used Xavier Initialization.
The source code of COCO-SBRS is available in \url{https://github.com/wzsong17/COCO-SBRS}.

\subsection{Recommendation Performance Evaluation and Analysis}
In this section, we conduct comparison experiments to evaluate the performance of the proposed COCO-SBRS and all baseline algorithms to answer the question: ``\textbf{How does the proposed COCO-SBRS perform compared with the baseline SBRSs?}''

Table \ref{table:exp1} presents the recommendation performance of all compared algorithms on three datasets.
In general, the first five SBRSs, i.e., SKNN, GRU4Rec, STAMP, CSRM and SR-GNN, do not perform well compared with other SBRSs, which consider the outer-session causes.
SKNN does not consider the complex item patterns and user preferences when calculating session similarity.
STAMP uses attention neural networks to model sessions and is good at extracting information from sessions with uninformative items. 
CSRM is a collaborative filtering-based SBRS, but its performance relies on the session similarities learned with attention networks and the quality of session embedding obtained by the memory neural network. 
GRU4Rec considers the sequential pattern among the items in sessions, while SR-GNN models the item transition patterns in sessions. 
Thus, their performance depends on the proportion of the corresponding patterns present in the dataset.

Different from the above five algorithms, HGRU, II-RNN, SASRec, BERT4Rec, and INSERT consider both inner-session causes and outer-session causes, i.e., user preference across sessions.
HGRU considers the dynamic change of user preferences between successive sessions of the same user and models it with a user-level RNN. 
In contrast, II-RNN considers that user preferences are constant between successive sessions of the same user, so II-RNN directly uses the information of the most recent session to supplement the limited information in the target session.
For SASRec and BERT4Rec, we concatenate the target session and all the historical interactions of the user as the interaction sequence, which allows the models to fully extract information from both the ISCs and OSCs based on their deep neural networks. 
We can see that BERT4Rec can obtain good performance on Reddit due to many repeat items in user sequences. 
INSERT is the state-of-the-art SBRS for next item recommendations in short sessions, which uses a specially designed session similarity calculation module to incorporate information from other users.
From Table \ref{table:exp1}, we can see that INSERT performs better than the above methods w.r.t. most of metrics.
However, INSERT still needs to find similar sessions from the training set. It has difficulty finding similar sessions from the data when considering both inner-session causes and outer-session causes.

The proposed COCO-SBRS achieves a significant improvement in both Recall and NDCG compared to the baseline methods (except for the NDCG in Reddit).
The reason is that the counterfactual computing framework alleviates the difficulty of finding similar sessions due to data sparsity while considering both inner-session causes and outer-session causes.
Besides, we pre-train BRM in the framework so that the model can better untangle and identify the true causes and model the causalities in SBRSs.
Finally, the performance of COCO is further enhanced by the boost factor.

\subsection{Ablation Analysis}
In this experiment, we test the performance of several simplified versions of COCO-SBRS to answer the question:
``\textbf{How does the proposed counterfactual computing framework benefit the next item recommendation in SBRS?}''

We design three simplified variants of COCO-SBRS: (1) \textbf{BRM}, the base recommendation model predicts the next item in sessions without the counterfactual computing framework; (2) \textbf{COCO w/o BF}, which removes the boost factor by setting $\epsilon=0$; (3) \textbf{COCO w/o SSL}, which removes the self-supervised loss, i.e., Equation (\ref{eq:l2}).

Table \ref{table:exp_ab} shows the performance of the three simplified variants and the full version of COCO-SBRS on Delicious and Lastfm.
The table shows that BRM performs the worst, showing the effectiveness of the proposed collaborative filtering-based counterfactual framework proposed in this paper. 
It also shows that even considering all the causes for the selection of the next item, it is difficult to model the causalities in SBRSs well using deep learning models only.
The poor performance of COCO w/o BF and COCO w/o SSL compared to the full version COCO-SBRS indicates that both the boost factor and the pre-training with self-supervised loss can effectively improve the performance of the proposed COCO-SBRS. 
The booster factor can improve the performance of COCO-SBRS by emphasizing the most recently seen items since, in real-world data, users often interact with preferred items repeatedly \cite{latifi2021session}. 
In addition, the self-supervised loss can help COCO-SBRS learn untangled representations of ISCs and OSCs and improve the ability to identify the true causes in Equation (\ref{eq:causalattn}).

\begin{table}[]
\centering
\caption{Recommendation performance of COCO-SBRS and its simplified variants on Delicious and Lastfm.}
\label{table:exp_ab}
\begin{tabular}{llcccc}
\hline
dataset                    & Variant      & R@5   & N@5   & R@20  & N@20  \\ \hline
\multirow{4}{*}{Lastfm}    & BRM          & 0.325 & 0.201 & 0.643 & 0.292 \\
                           & COCO w/o BF  & 0.444 & 0.253 & 0.727 & 0.335 \\
                           & COCO w/o SSL & 0.412 & 0.247 & 0.756 & 0.346 \\
                           & COCO-SBRS    & \textbf{0.504} & \textbf{0.289} & \textbf{0.793} & \textbf{0.374} \\ \hline
\multirow{4}{*}{Delicious} & BRM          & 0.214 & 0.137 & 0.397 & 0.190 \\
                           & COCO w/o BF  & 0.271 & 0.169 & 0.458 & 0.223 \\
                           & COCO w/o SSL & 0.273 & 0.170 & 0.463 & 0.225 \\
                           & COCO-SBRS    & \textbf{0.359} & \textbf{0.215} & \textbf{0.520} & \textbf{0.263} \\ \hline
\end{tabular}
\end{table}

\section{Conclusion}
The present study was designed to address the problem that the confounder in SBRSs can cause the SBRS models to learn spurious correlations in the data.
This work proposes a counterfactual-based framework named COCO-SBRS for next item recommendation in SBRSs.
COCO-SBRS first adopts a self-supervised approach to pre-train a recommendation model to learn the causalities in SBRSs,
and then make recommendations for a session based on some neighbor sessions with the same causes as this session and the recommendation model used to simulate the user's selection in sessions.
We conduct extensive experiments on three real-world datasets, and the results show that the proposed COCO-SBRS can eliminate the influence of spurious correlations caused by the confounder in SBRSs and make accurate next item recommendations.


\begin{acks}
This work is supported by NSFC (under Grant No.62202200, 61976050, 61972384), the Fundamental Research Funds for the Central Universities 2412019ZD013, and Jilin Science and Technology Department 20200201280JC.
\end{acks}
\bibliographystyle{ACM-Reference-Format}
\bibliography{ref}

\clearpage

\appendix

\section{Hyper-parameters Sensitivity Test}

\begin{figure}[h]
  \centering
  \includegraphics[width=1\linewidth]{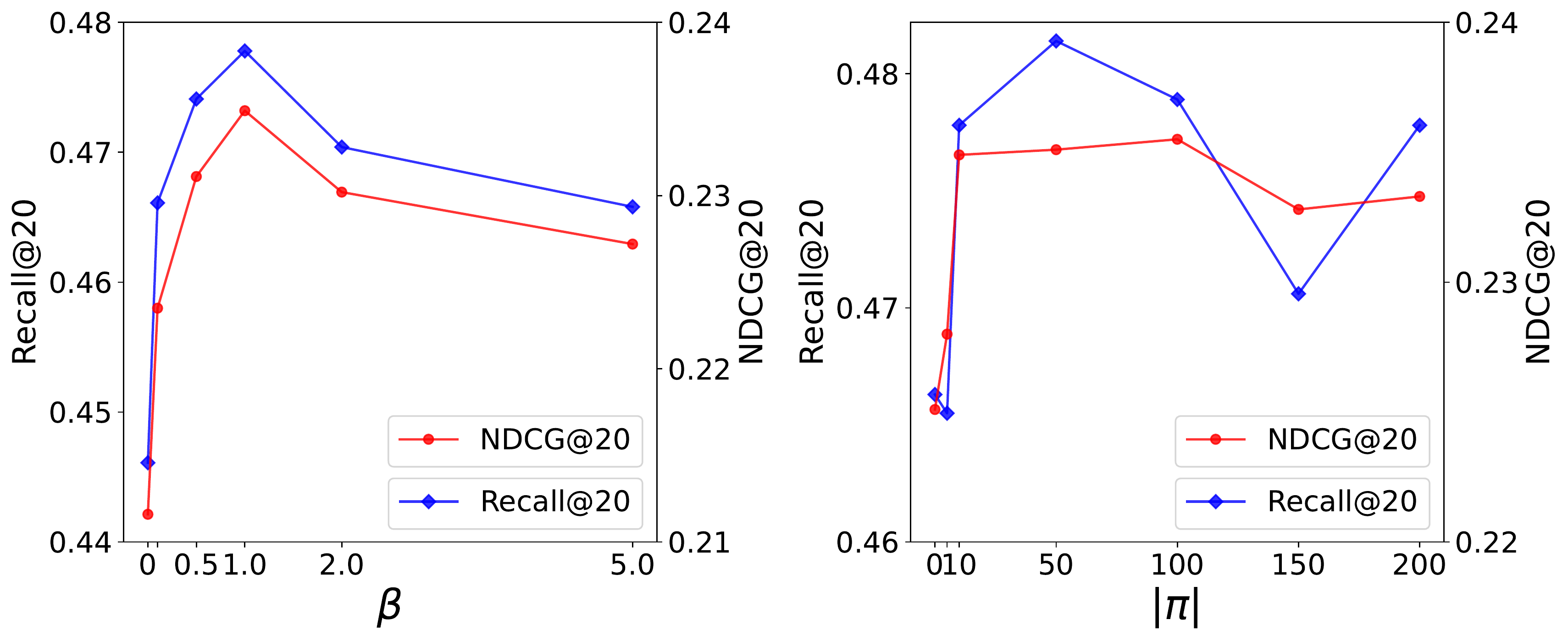}
  \caption{Sensitivity of $\beta$ and $|\pi|$ on Delicious.}
  \Description[Sensitivity of $\beta$ and $|\pi|$ on Delicious.] {COCO-SBRS performs best when beta=1 and |pai|=50}
  \label{fig:beta}
\end{figure}

We use two groups of experiments to test COCO's hyper-parameter sensitivity: (1) The process of pre-training BRM in COCO-SBRS considers two loss functions, i.e., Equation (\ref{eq:cel}) and Equation (\ref{eq:l2}), so in the first group of experiments we test the sensitivity of the model to the balance of two losses. 
(2) In collaborative filtering-based models, an important hyper-parameter affecting the algorithms' performance is the number of similar neighbors. In COCO-SBRS, this parameter refers to the number of other sessions that answer the counterfactual question, i.e., $|\pi|$. Thus, in the second group of experiments, we test the performance of COCO-SBRS under different $|\pi|$ while keeping other hyper-parameters unchanged.
The experimental results of the two groups of experiments are shown in Figure \ref{fig:beta}.

From Figure \ref{fig:beta}, we can see that the performance of COCO-SBRS increases as $\beta$ goes from 0 to 1 but gradually decreases after the $\beta$ is greater than 1. The experimental results indicate that the optimal trade-off weight for balancing the two losses is 1. Besides, the model performs best when $|\pi|=50$ but decreases as $|\pi|$ continues to increase. Note that for COCO-SBRS a large $|\pi|$ will result in more running time of the model, so in the previous experiments, we set $|\pi|=10$ to balance the recommended performance and model efficiency. 

\section{Recommendation Performance of All Methods on Sessions with Different Lengths}
\label{app_a}

\begin{figure}[htbp]
  \centering
  \includegraphics[width=1\linewidth]{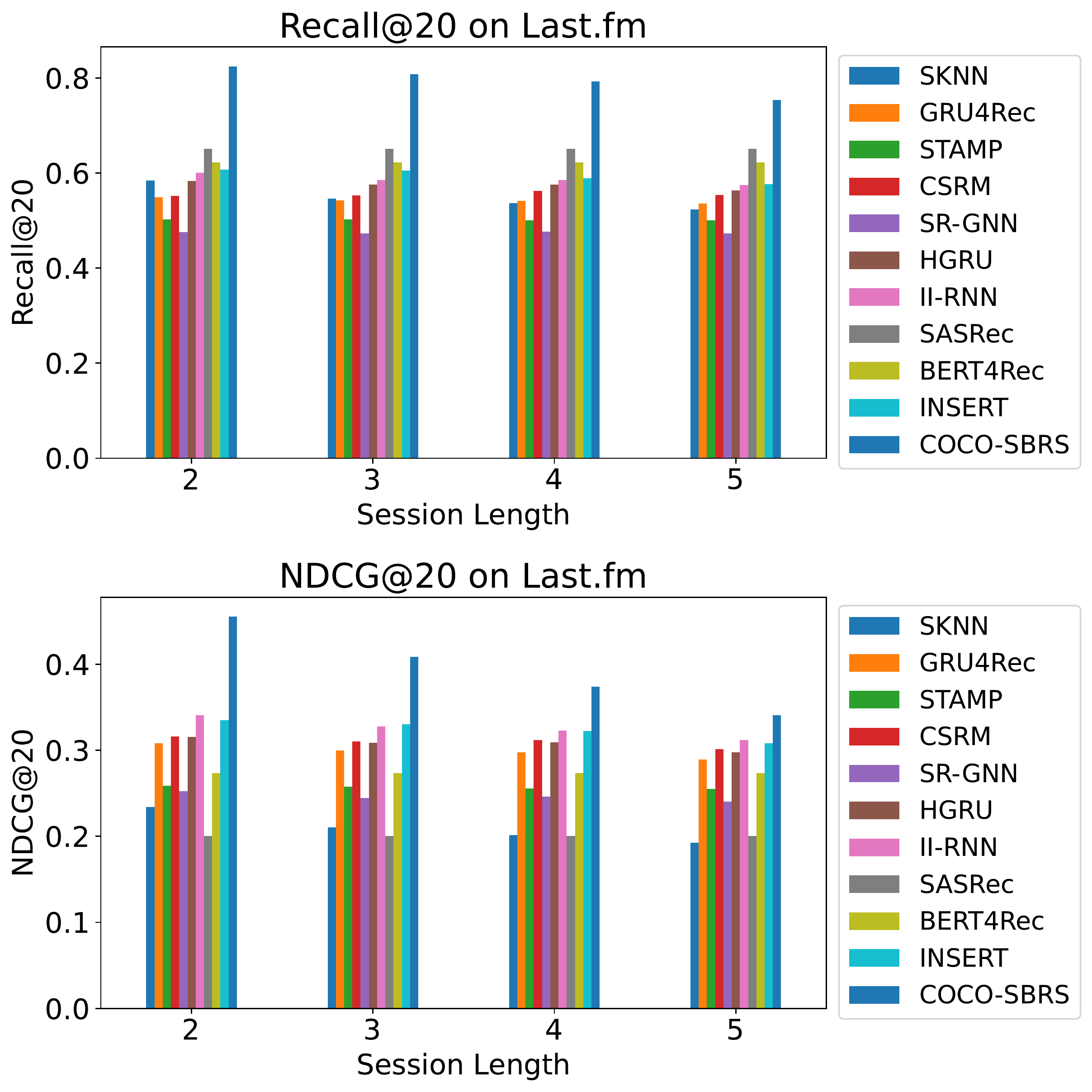}
  \caption{Recommendation Performance of All Methods on Sessions with Different Lengths on Last.fm. }
  \Description[Recommendation performance of all methods on sessions with different lengths on Last.fm.] {COCO-SBRS outperforms all baseline algorithms on sessions with all lengths.}
  \label{fig:length_l}
\end{figure}

\begin{figure}[htbp]
  \centering
  \includegraphics[width=1\linewidth]{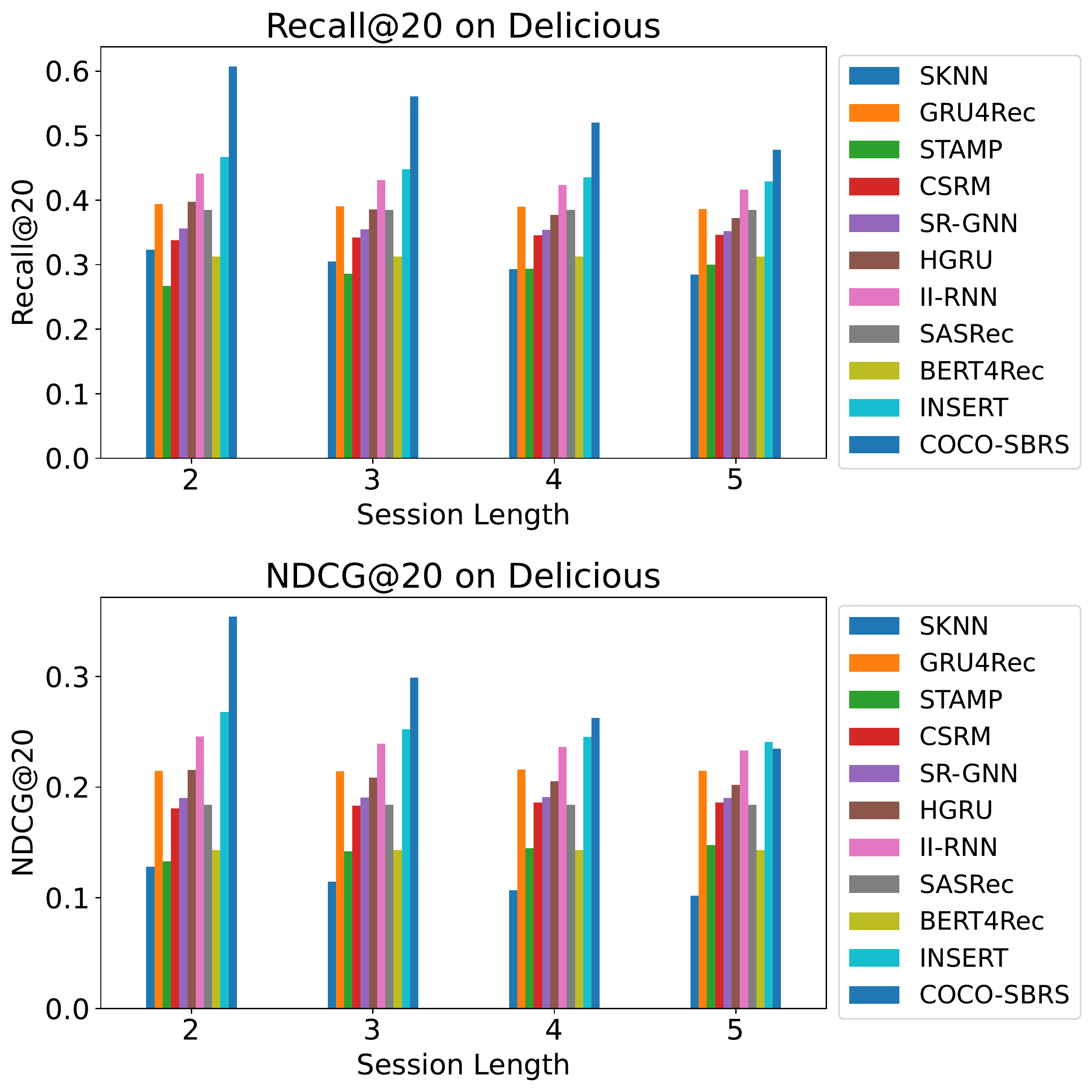}
  \caption{Recommendation Performance of All Methods on Sessions with Different Lengths on Delicious. }
  \Description[Recommendation preformance of all methods on sessions with different lengths on Delicious.] {COCO-SBRS outperforms all baseline algorithms on sessions with all lengths.}
  \label{fig:length_d}
\end{figure}

\begin{figure}[htbp]
  \centering
  \includegraphics[width=1\linewidth]{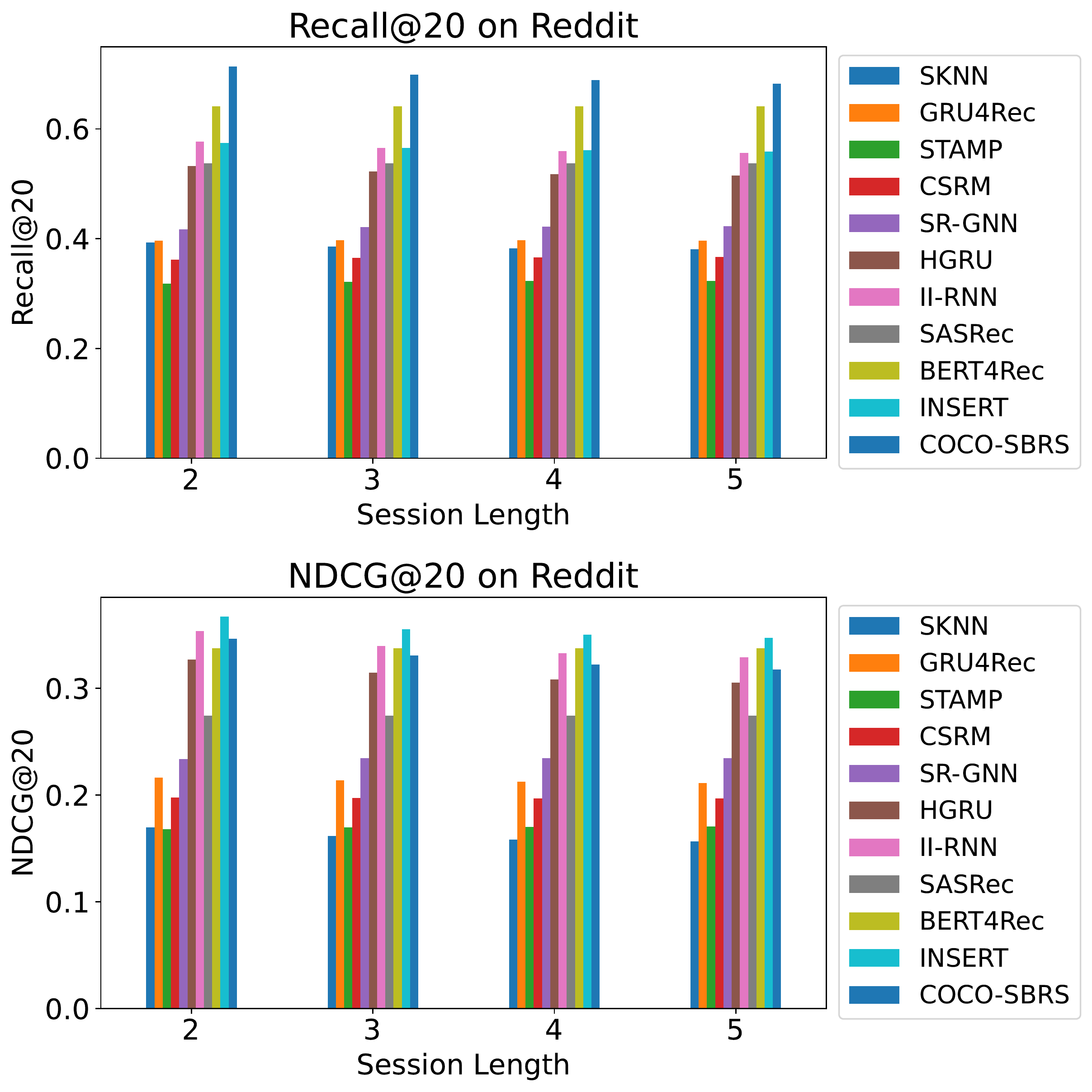}
  \caption{Recommendation Performance of All Methods on Sessions with Different Lengths on Reddit. }
  \Description[Recommendation preformance of all methods on sessions with different lengths on Reddit.] {COCO-SBRS outperforms all baseline algorithms on sessions with all lengths.}
  \label{fig:length_r}
\end{figure}

In this section, we test the performance of all compared methods on sessions with different lengths.
The length of a session refers to the number of interactions contained in the session \cite{insert}. The shorter the session length, the less information contained in the context of the session, and the lower the probability that the user select an item because of the ISCs, i.e., the information in the session context.
Therefore, the performance of each algorithm on sessions with different lengths can reflect their ability to learn the causality and the correlation among ISCs, OSCs and the interactions in SBRS.
We conduct comparison experiments on three datasets, and the results are depicted in Figure \ref{fig:length_l}, Figure \ref{fig:length_d} and Figure \ref{fig:length_r}.

From the figures, we can see that:
\textbf{i)} On all three datasets, COCO-SBRS outperforms all baseline algorithms in terms of Recall@20 on sessions with all lengths. COCO-SBRS also has the best NDCG@20 on Last.fm and Delicious, except for sessions of length five on Delicious.
This result shows that COCO-SBRS can better model the causalities in SBRS and avoid the model learning spurious correlations in the data. 
\textbf{ii)} There is a trend of decreasing performance of COCO-SBRS when session lengths become longer.
COCO-SBRS performs better than all baselines when the the session length is not greater than five.
This shows that COCO-SBRS is not good at extracting information from long sessions, which may be because we adopt a simple attention neural network to model the ISCs for sessions. 
Thus, we suggest that for those datasets containing a large number of long sessions, it's better to replace the attention-based model in COCO-SBRS with a more powerful session encoder to model ISCs in SBRSs.
\textbf{iii)} On Reddit, the methods consider both ISCs and OSCs, i.e., HGRU, II-RNN, SASRec, BERT4Rec, INSERT and COCO-SBRS, perform better than those methods that consider ISCs only.
This is because the preferred topics of each user, i.e., the items on Reddit, are stable, so users' long-term static preferences, i.e., OSCs, are the main causes for users to select the next topic. Thus, it is more difficult to predict user-item interactions based on the co-occurrence and sequential patterns of the topic, i.e., ISCs.

\clearpage



\end{document}